\documentclass[%
preprint,
amsmath,amssymb,
aps,
prb,
floatfix,
]{revtex4-1}

\usepackage{amsmath}    
\usepackage{mathtools}
\usepackage{siunitx}

\usepackage{graphicx}   
\usepackage{verbatim}   
\usepackage{color}      
\usepackage{subfigure}  
\usepackage{hyperref}   
\begin{document}

\title{Highly accurate and efficient algorithm for electrostatic interaction of charged particles
confined by metallic parallel plates}
\author{Samare Rostami, S. Alireza Ghasemi$^{*}$ and Ehsan Nedaaee Oskoee}
\affiliation{$^1$ Institute for Advanced Studies in Basic Sciences, P.O. Box 45195-1159, Zanjan, Iran}

\email[]{aghasemi@iasbs.ac.ir}

\date{\today}

\begin{abstract}
We present an accurate and efficient algorithm to calculate the electrostatic interaction of
charged point particles with
partially periodic boundary conditions that are confined along the nonperiodic
direction by two metallic parallel plates.
The method preserves the original boundary conditions and hence,
it does not introduce any kind of artifacts.
In addition, it enjoys the quasilinear complexity of $\mathcal{O}(N\ln(N))$, where $N$
being the number of particles in the simulation box.
In fact, based on the superposition principle in electrostatics, the problem is split into two
electrostatic problems where each one can be calculated by the appropriate Poisson solver.
In this paper we apply the method to sodium chloride ultrathin films and investigate
its dielectric response with respect to external bias voltage.
We show how accurately in this method one can obtain the total charge induced on metallic
boundaries to an arbitrary precision.
\end{abstract}

\maketitle

\section{Introduction\label{sec:introduction}}
In many experiments in physics and chemistry, an ionic material with slablike
geometry is sandwiched between two metallic parallel plates
with electric potential difference.
This situation occurs
in solid oxide fuel cells and batteries where an electrolyte
lies between two electrodes as well as
dielectric materials in capacitors.
Atomistic simulations are of great importance
to improve the understanding of the atomic scale mechanisms and processes
in ionic materials used as an electrolyte in fuel cells or dielectric in capacitors.
Electrostatic interaction in ionic materials dominates among interatomic forces
and its proper calculation is thus very important for the simulations of such materials.
In atomistic simulation of electrolytes,
the two electrodes are usually neglected since computational methods
known for Coulombic interaction
do not account for the effects of the electrodes.
Materials with slablike geometries are modelled with periodic
boundary conditions for lateral directions and free
boundary condition for the perpendicular direction.
This type of boundary conditions is typically called
surface, slab, or 2D+h boundary conditions.
The electrostatic interaction between charged particles and
the two electrodes are typically modelled by an external electric
field in the nonperiodic direction.
Such a modelling does not provide information on
various physical properties such as
the charge density induced at the electrodes and
the microscopic response of particles in the vicinity of the
interface between the electrolyte and the electrodes.
In addition, the effect of a uniform external electric field
may be different from the electrostatic interaction
in the electrolyte with the presence of 
the two electrodes, in particular when the electrolyte is ultrathin.
In addition, there may be nonlinear phenomena that cannot be investigated
by using a uniform electric field.

Due to the long ranged nature of the Coulomb potential,
direct summation for free boundary conditions scales as $\mathcal{O}(N^2)$,
where $N$ is the number of particles in the simulation box.
Several methods~\cite{Greengard1987,Neelov2007} have been developed
for the calculation of Coulomb interaction with free boundary conditions that
have complexity of $\mathcal{O}(N\ln(N))$ while preserving the
original boundary conditions.
For periodic boundary conditions, particles in the
periodic images must be considered as well.
The Ewald method~\cite{Ewald1921} for fully periodic boundary conditions
solves the problem due to the inclusion of particles in periodic images
in the calculations, however, it has time complexity of
$\mathcal{O}(N^2)$ and by choosing optimal parameters
it can be improved to $\mathcal{O}(N^\frac{3}{2})$ scaling.
The scaling can be enhanced further to an $\mathcal{O}(N\ln(N))$ if
fast Fourier transformation is employed.
Several methods are developed for Coulombic interaction with slab
boundary conditions.
Some of these methods are in the spirit of Ewald
methods~\cite{Ghasemi2007,ELC1,ELC2,EWALD2D1,EWALD2D2,EWALD2D3,Spohr,Rhee,Yeh}
and the others~\cite{MMM2D}
are based on the MMM~\cite{Sperb} method.
The MMM method for slab boundary conditions is called
MMM2D~\cite{MMM2D} and it has the advantage that it preserves the original
boundary conditions, however, its complexity is  $\mathcal{O}(N^\frac{5}{3})$.
The standard implementation of the Ewald method based on
fast Fourier transformation have been used in atomistic simulation
of charged particles in slablike geometries.
The major problem of the standard Ewald method for slab boundary conditions
is due to the assumption that boundary conditions in all directions
are considered as periodic.
Therefore, in the nonperiodic direction an empty space must be
used to decouple the interaction between the particles in the simulation
box and those of periodic images.
This approach not only makes the method inefficient but also
the artificial interaction does not vanish fast enough with the
size of the empty space if the system has dipole moment.
Methods such particle particle particle mesh with layer correction~\cite{ELC1,ELC2} (P3MLC)
have been proposed to correct for dipole-dipole interaction between periodic images.
In 2007, a new method was introduced by Ghasemi
and coworkers that was called
particle particle particle density (P$^3$D) method~\cite{Ghasemi2007}.
The P$^3$D method is in the spirit of Ewald method and it employs
plane waves for the periodic directions and
finite element for the nonperiodic direction
to expand the potential function and the charge density.
The P$^3$D method is the only method which
not only preserves the original slab boundary conditions
but also it has desirable scaling of $\mathcal{O}(N\ln(N))$.

None of the aforementioned methods in their original form
can be directly employed for the electrostatic calculation
of charged particles confined by metallic boundaries.
In recent years, two methods were developed by
Holm~\cite{ICMMM2D,ICC*1}
and coworkers that compute electrostatic interaction of
charged particles confined between two dielectric materials.
The image charge MMM2D~\cite{ICMMM2D}(ICMMM2D) is based on
the method of image charge in electrostatics and the MMM2D method.
Even though it was originally developed for Coulombic interaction
of charged particles inside a dielectric material trapped between two
dielectric materials,
it can be applied to electrostatic interaction of charged particles
confined by metallic plates.
In ICMMM2D, the number of image charge particles strongly depends on
the difference between dielectric constants of the electrolyte
and those of surrounding materials.
The computational cost of including image charge particles
scales linearly with the number of particles in the simulation box.
However, for large number of particles the original
scaling of MMM2D is preserved.
In principle, the method can also be applied for Coulombic interaction
of charged particles between metallic plates.
Based on induced charge computation~\cite{Boda2004}
(ICC) method, Tyagi~\cite{ICC*1} 
and coworkers introduced a generic method
that can be used for dielectric interface of arbitrary shape.
The method requires a Poisson solver suitable for
the same geometry in the absence of the dielectric interface.
The method is called
ICC$^*$~\cite{ICC*1} in which the star stands
as a representative for any Poisson solver one employs.
The ICC$^*$ algorithm scales basically like the Poisson
solver employed in the method.
In 2013 Takae and Onuki~\cite{Takae2013}
introduced a method for electrostatic interaction of
charged and polar particles between metallic plates.
The method is an extension of the Ewald method which
accounts for the effect of image particles.

In this paper, we present a method to calculate
electrostatic interaction of charged point particles between
two parallel metallic plates.
The method is based on superposition principle
in electrostatics.
In fact, similiar approaches can be applied to electrostatic
interaction of charged point particles or continuous charge densities
trapped by metallic boundaries with arbitrary shape.
In this paper, we focus on slab geometry and we show the particular
advantage of the method for this type of geometry.
The method scales basically like the Poisson
solver of the problem in the absence of the metallic plates.
For slab geometry we employ the P$^3$D
method~\cite{Ghasemi2007} and the algorithm scales as
$\mathcal{O}(N\ln(N))$.
Furthermore, the method is accurate and the error in the electrostic
energy and other physical quantities of interest such as total charge induced on
one of the electrodes can be reduced to an arbitray value. 


\section{Method}
\begin{figure*}
\centering
\includegraphics[width=0.99\textwidth]{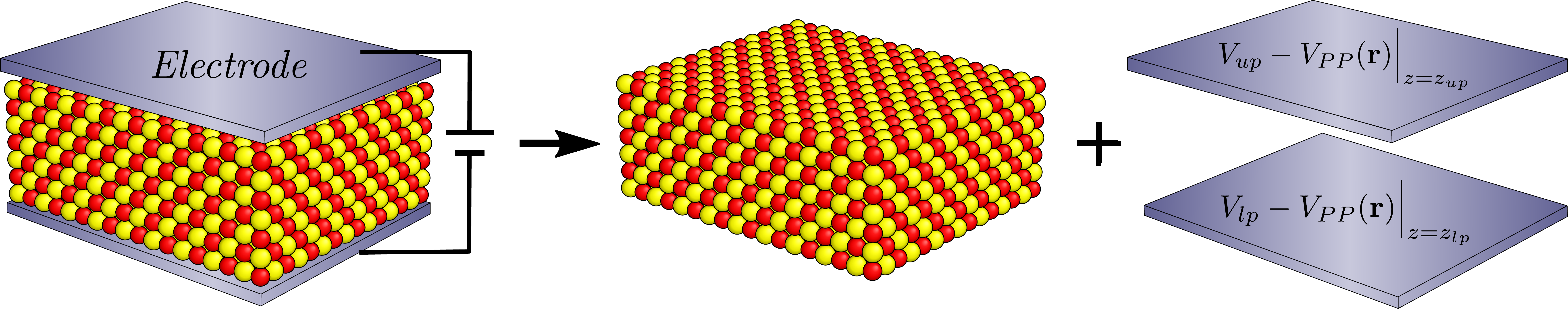}
\caption{(color online) Schematic illustration of the method where the electrostatic interaction of
charged point particles confined by metallic parallel plates (left) is split
into two parts; a system of charged point particles
with free boundary condition in the $z$ direction and periodic
in $x$ and $y$ directions, (middle)
and two metallic parallel plates with boundary conditions
given in Eq.~(\ref{eqn:superposition}) (right).
\label{fig:p3dbias_schematic}}
\end{figure*}

\subsection{Splitting the electrostatic problem based on
superposition principle\label{sec:superposition}}
Consider a system of $N$ particles with charges $q_i$ at
positions ${\bf r}_i$ in an overall neutral and rectangular
simulation box of dimensions $L_x$, $L_y$, and $L_z$ with
periodic boundary conditions in $x$ and $y$ directions.
In order to model the electrodes,
boundary conditions at $z=0$ and $z=L_z$ are at constant voltages
at $V_{lp}$ and $V_{up}$, respectively.
In fact, $\Delta V=V_{up}-V_{lp}$ indicates the potential different between the two electrodes.
The electrostatic interaction of charged particles with such
boundary conditions is not Coulombic because $\frac{1}{r}$ function
does not satisfy boundary conditions at $z=0$ and $z=L_z$.
The electric potential must fulfil the Poisson's equation,
\begin{align}\label{eqn:poisson}
\nabla^2\,V({\bf r})=-4\pi\rho({\bf r}),
\end{align}
where $\rho({\bf r})$ is the charge density of point particles.
The Green function of Poisson's equation for slab geometry with
metallic boundary conditions is complicated and no compact
functional form of the Green function is known.

We propose a method based on the superposition principle
in electrostatics that can be used for the calculation of
electrostatic interaction of charged particles or continuous
charge densities surrounded with metallic materials.
The superposition principle in electrostatics, which is due to the
fact that Poisson's equation is a linear differential equation,
allows us to split the problem into two problems.
The corresponding electric potentials of the two problems, called
$V_1({\bf r})$ and $V_2({\bf r})$, fulfil Poisson's equation with different
charge densities, $\rho_1({\bf r})$ and $\rho_2({\bf r})$, respectively.
Boundary conditions of the two problems are different as well.
Both the new charge densities and boundary conditions are
arbitrary and can be anything provided $\rho({\bf r})=\rho_1({\bf r})+\rho_2({\bf r})$
and the two boundary conditions sum up to the original boundary conditions.
Here we choose $\rho_1({\bf r})=\rho({\bf r})$ and $\rho_2({\bf r})=0$ while
for slab geometries the boundary conditions of the first problem is taken to be
free in $z$ direction and periodic in $x$ and $y$ directions.
In this way one can employ any Poisson solver for slablike geometry
that have been summarized in Sec.~\ref{sec:introduction}.
From now on $V_1({\bf r})$ and $V_2({\bf r})$ are replaced
by $V_{pp}({\bf r})$ and $V_s({\bf r})$, respectively,
to emphasize that the first problem is the
electric potential due to charged point particles and the second problem
is a smooth potential which does not diverge any point within the simulation box.
Poisson solver of the first problem must provide potential values
at boundaries.
Once $V_{pp}({\bf r})$ is calculated we can solve the Laplace equation of
the second problem with boundary conditions as
\begin{align}\label{eqn:superposition}
\left(V({\bf r})-V_{pp}({\bf r})\right)\Big|_{\text{boundaries}}
\end{align}

In our calculations for the Coulombic interaction in the first problem,
we employ the P$^3$D method which can provide potential values
at any point in the simulation box while it is both efficient and accurate.
It is also possible to use the method developed by Genovese~\cite{Genovese2007} and coworkers
to solve the first problem.
In the next section we reiterate a summary of the P$^3$D method.
The Laplace equation of the second problem is solved in a similar approach
used in the P$^3$D method.
For slablike geometry, Eq.~(\ref{eqn:superposition})
is applied only to boundary conditions in the nonperiodic direction, i.e. $z=0$ and $z=L_z$.
Fig.~\ref{fig:p3dbias_schematic} shows a schematic view of the method presented
in this section.

\subsection{P3d Method\label{sec:p3d}}
Consider a charge neutral set of $N$ charged point particles
with charges $q_i$ located at
position ${\bf r}_i$ in a $L_x \times L_y\times L_z$ simulation box.
Suppose that the particles are subjected to periodic boundary conditions
in two directions and free boundary conditions in the third
(here we let $z$ to be the nonperiodic direction).
The total electric potential energy of this systems is given by
\begin{align}
\label{eqn:coulomb_slab}
E=\frac{1}{2}\sum_{\bf n}' \sum_{i,j=1}^{N} \frac{q_i q_j}{|{\bf r}_{ij} + {\bf n}|}
\end{align}
where ${\bf r}_{ij}={\bf r}_{i}-{\bf r}_{j}$ and ${\bf n}=(n_x L_x,n_y L_y,0)$
with $n_x$ and $n_y$ being integers.
The prime on the outer sum denotes that for ${\bf n}=0$,
the term $i=j$ has to be excluded.
Ewald~\cite{Ewald1921} showed that Eq.~(\ref{eqn:coulomb_slab}) can be split into two parts
which one of them decays rapidly in real space and 
the other is a smooth charge density which can be treated in Fourier space very efficiently.
Following Ewald's approach we get,
\begin{align}
E&= \frac{1}{2}\sum_{\bf n}'\sum_{i,j=1}^{N}
\left[\frac{q_i q_j}{|{\bf r}_{ij} + {\bf n}|}
- \int \!\!\! \int \frac{\rho_i({\bf r}) \rho_j({\bf r'}+{\bf n})}
{|{\bf r}-{\bf r'}|}  d{\bf r}  d{\bf r'}\right]  \nonumber \\
&+\frac{1}{2} \sum_{\bf n}
\sum_{i,j=1}^{N}\int \!\!\! \int
\frac{\rho_i({\bf r}) \rho_j({\bf r'}+
 {\bf n})}{|{\bf r}-{\bf r'}|} d{\bf r}  d{\bf r'}  \nonumber \\
& -\frac{1}{2} \sum_{i=1}^{N}\int \!\!\! \int
\frac{\rho_i({\bf r}) \rho_i({\bf r'})}{|{\bf r}-{\bf r'}|} d{\bf r}  d{\bf r'}
\label{eqn:energy_pp}
\end{align}
where $\rho_i(r)$ is a smooth spherical charge densities
centered on the particle positions.
The standard choice in Ewald method for the smooth atomic charge density is 
a Gaussian function,
\begin{align}
\label{eqn:gaussian_charge}
\rho_i({\bf r})=\frac{q_i}{(\alpha^2 \pi)^{\frac{3}{2}}}
\; \; \exp\left[-\frac{|{\bf r}-{\bf r}_i|^2}{\alpha^2}\right].
\end{align}
Therefore, Eq.~(\ref{eqn:energy_pp}) can be rewritten as
\begin{align*}
E=& E_{short}+E_{long}-E_{self},
\end{align*}
where
\begin{subequations}
\begin{align}\label{eqn:energy_pp_a}
E_{short}=& \frac{1}{2}\sum_{\bf n}' \sum_{i,j=1}^{N}
\frac{q_i q_j \; \; \rm{erfc}\, \left[\frac{|{\bf r}_{ij}
+ {\bf n}|}{\alpha\sqrt{2}}\right]}
{|{\bf r}_{ij} + {\bf n}|} \\
E_{long}=& \frac{1}{2} \sum_{\bf n} \label{eqn:energy_pp_b}
\sum_{i,j=1}^{N}\int \!\!\! \int
\frac{\rho_i({\bf r}) \rho_j({\bf r'}+ {\bf n})}{|{\bf r}-{\bf r'}|}
d{\bf r}  d{\bf r'} \\
E_{self}=& \frac{1}{\alpha \sqrt{2 \pi}} \sum_{i=1}^{N} q_i^2. \label{eqn:energy_pp_c}
\end{align}
\end{subequations}
Complementary error function in Eq.~(\ref{eqn:energy_pp_a}) decays exponentially.
It can thus be made of finite range by introducing a cutoff.
Therefore, the calculation of the first term can be done with linear scaling.

The system is considered to  have a nonzero charge density just within  
$[z_{lb},z_{ub}]$ for nonperiodic direction. 
The simulation cell in periodic directions are the same size as the original
cell and in z direction, $z_{ub}-z_{lb}$ is $L_z$ plus twice the cutoff for
Gaussian charge density.
Hence we can write the domain as,
\begin{align}\label{eqn:cell_poisson}
\mathcal{V}:=[0,L_x]\otimes [0,L_y]\otimes [z_{lb},z_{ub}].
\end{align}
$E_{long}$ is the electrostatic energy of charge density comprised of
superposition of atomic Gaussian functions.
The electric potential due to atomic Gaussian functions, $V_{GF}(r)$,
can be obtained by solving Poisson's equation,
\begin{align}
\label{eq22}
\nabla^2 V_{GF}({\bf r})=-4\pi\rho({\bf r})
\end{align}
where $\rho({\bf r})=\sum_{i=1}^{N}\rho_i({\bf r})$.
Eq.~(\ref{eq22}) is solved for simulation box given in Eq.~(\ref{eqn:cell_poisson})
with periodic boundary conditions in $x$ and $y$ directions.
We expand the potential function and the charge density in terms of
Fourier series,
\begin{subequations}
\begin{align}
V_{GF}(x,y,z)&\hspace{-0.15cm}=\hspace{-0.25cm}&\sum_{k,l=-\infty}^{\infty}
c_{kl}(z) \exp\left[2i\pi(\frac{k \, x}{L_x}+\frac{l \, y}{L_y})\right],
\label{eqn:Fourier_pot} \\
\rho(x,y,z)&\hspace{-0.15cm}=\hspace{-0.25cm}&
\sum_{k,l=-\infty}^{\infty} \frac{\eta_{kl}(z)}{-4\pi}
\exp\left[2i\pi(\frac{k\,x}{L_x}+\frac{l\,y}{L_y})\right].
\label{eqn:Fourier_rho}
\end{align}
\end{subequations}
Inserting Eqs.~(\ref{eqn:Fourier_pot}) and (\ref{eqn:Fourier_rho}) in Eq.~(\ref{eq22}) yields,
\begin{align}
\label{eqn:poisson_z}
\left( \frac{d^2}{dz^2} - g_{kl}^2\right) c_{kl}(z) =\eta_{kl}(z), 
\end{align}
where
\begin{align}\label{eqn:g_kl}
g_{kl}:=2\pi\sqrt{\frac{k^2}{L_x^2}+\frac{l^2}{L_y^2}}
\end{align}
In order to calculate Fourier coefficients $c_{kl}(z)$ in Eq.~(\ref{eqn:poisson_z})
one needs to determine boundary conditions at $z\rightarrow \pm\infty$.
As it is explained in Ref.~[\onlinecite{Ghasemi2007}],
we have $ V(x,y,z\rightarrow \pm\infty) = \mp\beta$ 
where $\beta$ is proportional to the dipole moment of the charge distribution along the z direction,
\begin{align}
\beta= \frac{1}{2} \int_{z_{lb}}^{z_{ub}}\eta_{00}(z') z' dz'.
\label{eq15}
\end{align}
To deal with the Eq.~(\ref{eqn:poisson_z}) with above boundary conditions, one can
write the following conditions for the $g$'s :
\begin{itemize}
\item $g_{00}=0 \Rightarrow \frac{d^2}{dz^2} c_{00}(z) =\eta_{00}(z)$, 
we solve this differential equation with boundary condition
$c_{00}(z\rightarrow \pm\infty)=\mp \beta$
\item $g_{kl}\neq0 \Rightarrow \left( \frac{d^2}{dz^2} - g_{kl}^2\right) c_{kl}(z) =\eta_{kl}(z)$,
for all of these differential equations we have to impose boundary conditions of the form
$c_{kl}(z\rightarrow\pm\infty)=0$.
\end{itemize}
$c_{00}(z)$ is constant for $z\notin[z_{lb},z_{ub}]$ and therefore,
one has to impose the Dirichlet boundary conditions
$c_{00}(z_{lb})=\beta$ and $c_{00}(z_{ub})=-\beta$.
It is explained in Ref.~[\onlinecite{Ghasemi2007}] that
all differential equations with $|k|+|l|>0$ must be solved subject to Robin
boundary conditions given by the following equations at $z_l$
\begin{align}
\label{eq17_BC1}
c'(z_{lb}) - g_{kl} c(z_{lb})=0,
\end{align}
and at $z_{ub}$
\begin{align}
\label{eq18_BC3}
c'(z_{ub}) + g_{kl} c(z_{ub})=0.
\end{align}
To solve the differential equations in Eq.~(\ref{eqn:poisson_z}) with boundary conditions explained above,
finite element method are employed as explained in Ref.~[\onlinecite{Ghasemi2007}] and its appendix.
Once $c_{kl}(z)$ are obtained, the potential function is calculated using a reverse Fourier transform.
Then the electrostatic energy $E_{long}$ and atomic forces can be calculated.
The P$^3$D method provides values of the potential function at upper and lower planes that
are necessary to build the boundary conditions of the second problem.


\subsection{Laplace's equation of the second problem}
As it is explained in sec.~\ref{sec:superposition}, we should calculated
values of the potential function of the first problem at boundaries.
The potential function of the P$^3$D method is given by
\begin{align}\label{eqn:pot_p3d}
V_{pp}({\bf r})=V_{GF}({\bf r})+\sum_{\bf n} \sum_{i=1}^{N}
\frac{q_i \; \; \rm{erfc}\, \left[\frac{|{\bf r}_{i}-{\bf r}+
{\bf n}|}{\alpha}\right]}
{|{\bf r}_{i} - {\bf r} + {\bf n}|}.
\end{align}
The first term in Eq.~(\ref{eqn:pot_p3d}) is available on the grid points.
If the grid points are chosen such that the upper and lower boundaries
lie on the grid points, the first term can be calculated on boundaries at no cost
and without interpolation, as it is done in our implementation.
The second term in Eq.~(\ref{eqn:pot_p3d}) decays exponentially
and it can be made of finite range by introducing a cutoff.
The Laplace equation must be solved subject to the following
boundary conditions for the upper plane
\begin{align}\label{eqn:bc_up_laplace}
V_{bup}(x,y)=V_{up}-V_{pp}({\bf r})\Big|_{z_{up}},
\end{align}
and for the lower plane
\begin{align}\label{eqn:bc_lp_laplace}
V_{blp}(x,y)=V_{lp}-V_{pp}({\bf r})\Big|_{z_{lp}},
\end{align}
where in our implementation $z$ coordinate of lower and upper planes,
$z_{lp}$ and $z_{up}$, are zero and $L_z$, respectively.
In a similar approach to sec.~\ref{sec:p3d},
we expand the potential function in terms of Fourier series for $x$ and $y$ directions
with expansion coefficients of $f_{kl}(z)$.
Replacing the Fourier series expansion in the Laplace equation, it yields
\begin{align}\label{eqn:laplace_z}
\left( \frac{d^2}{dz^2} - g_{kl}^2\right) f_{kl}(z) =0, 
\end{align}
where $g_{kl}$ is given by Eq.~(\ref{eqn:g_kl}).
Eq.~(\ref{eqn:laplace_z}) differs from Eq.~(\ref{eqn:poisson_z}) in
two aspects, first it is homogeneous and second its boundary conditions is different.
In order to obtain boundary conditions of Eq.~(\ref{eqn:laplace_z}),
$V_{bup}(x,y)$ and $V_{blp}(x,y)$
are expanded in term of Fourier series for $x$ and $y$ directions
with expansion coefficients of $a_{kl}$ and $b_{kl}$, respectively.
Eq.~(\ref{eqn:laplace_z}) can be solved analytically and there is no need to
employ the finite element method used for solving Eq.~(\ref{eqn:poisson_z}).
$f_{00}(z)$ is a linear function given by
\begin{align}\label{eqn:laplace_sol_00}
f_{00}(z)=\frac{(a_{00}-b_{00})\,z+(b_{00}\,z_{up}-a_{00}\,z_{lp})}{z_{up}-z_{lp}},
\end{align}
and for $f_{kl}(z)$ with $|k|+|l|>0$, we have
\begin{align}\label{eqn:laplace_sol_kl}
f_{kl}(z)&= \frac{a_{kl}\,\sinh(g_{kl}(z-z_{lp}))+b_{kl}
\sinh(g_{kl}(z_{up}-z))}{\sinh(g_{kl}(z_{up}-z_{lp}))}.
\end{align}
Using Eqs.~(\ref{eqn:laplace_sol_00}) and (\ref{eqn:laplace_sol_kl}),
$f_{kl}(z)$ are calculated on the grid points and
subsequently $V_s({\bf r})$ are obtained on the grid points
by performing a reverse Fourier transformation.
The electrostatic energy contribution of the
second problem is given by
\begin{align}\label{eqn:es_energy_smooth}
E_s=\sum_{i=1}^{N} q_i V({\bf r}_i).
\end{align}
The potential function, $V_s({\bf r})$, is not known at atomic positions.
Therefore, we use Lagrange polynomial of order eight to interpolate the potential
function at atomic positions.
The total electrostatic energy is the sum of the four terms,
three of which are due to the electrostatic energy of point particles
given by Eqs.~(\ref{eqn:energy_pp_a}), (\ref{eqn:energy_pp_b}), (\ref{eqn:energy_pp_c})
and the fourth term is the electrostatic energy of the smooth potential given by
Eq.~(\ref{eqn:es_energy_smooth}).

\section{Numerical Results}

\begin{figure}
\centering
\includegraphics[width=0.45\textwidth]{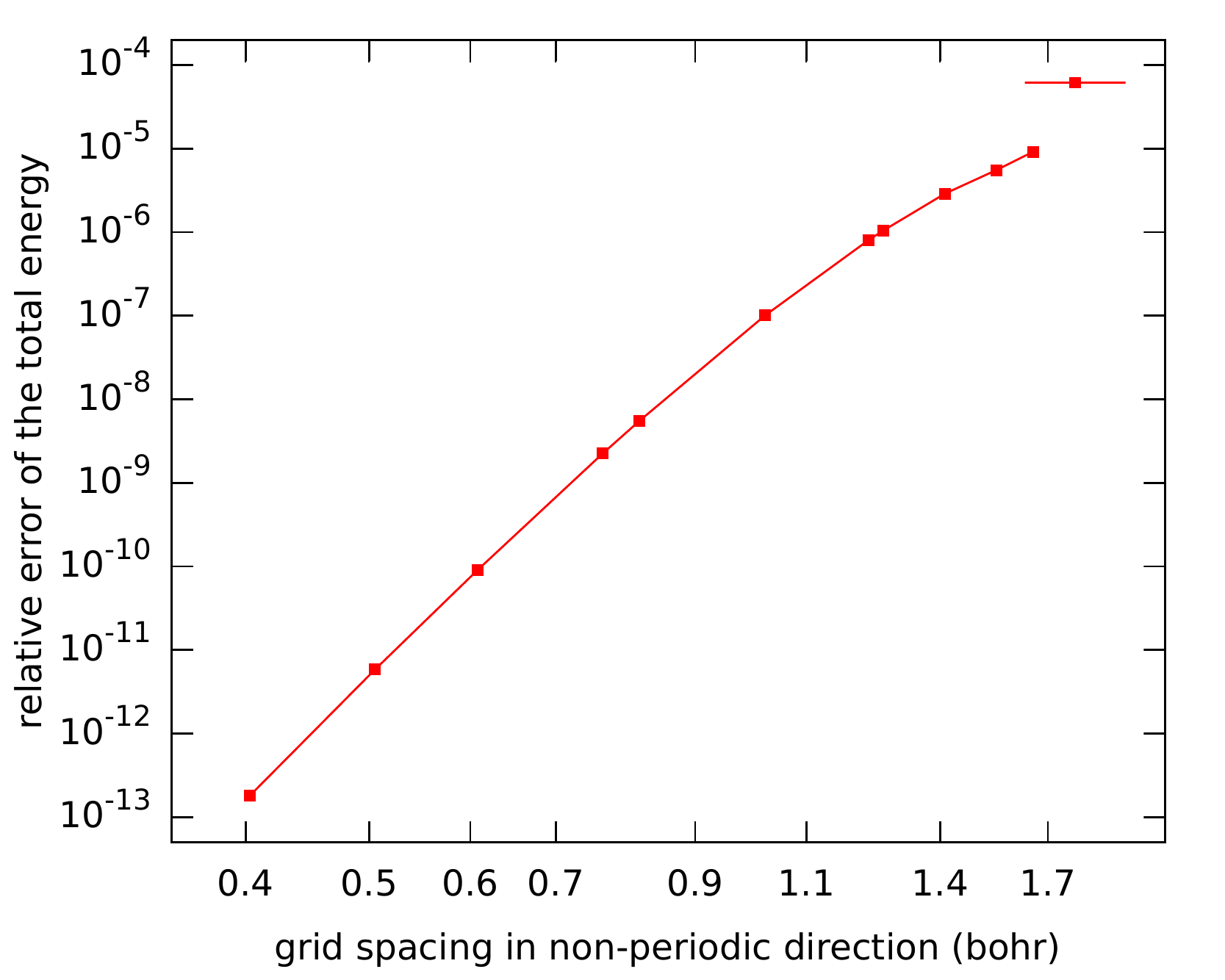}
\caption{(color online) Relative error in the total energy versus grid spacing in $z$ direction.
On this double logarithmic plot, the curve has an asymptotic
slope of $14$ and machine precision can be reached.
\label{fig:epoterr}}
\end{figure}
We illustrate the accuracy of the method for two important quantities
relevant to electrostatic problems of point particles confined by metallic boundaries.
A fast and robust convergence in electrostatic energy is vital for an
algorithm for solving Poisson's equation.
Due to the use of Fourier series in periodic directions, exponential
convergence rate is observed with respect to grid spacing in those directions.
However, the convergence rate in the nonperiodic direction, due to the use of
polynomial as the basis set, is algebraic $\mathcal{O}(h_z^{2m})$
where $m$ is the degree of the polynomial used in the finite element method.
Fig.~\ref{fig:epoterr} illustrates the convergence rate in total energy
in terms of grid spacing in the nonperiodic direction.
Detailed explanation of the finite element approach employed in this work is
given in Ref.~[\onlinecite{Ghasemi2007}].
Fig.~\ref{fig:epoterr} shows that the error in total energy can be
reduced down to machine precision.
\begin{figure}
\centering
\includegraphics[width=0.45\textwidth]{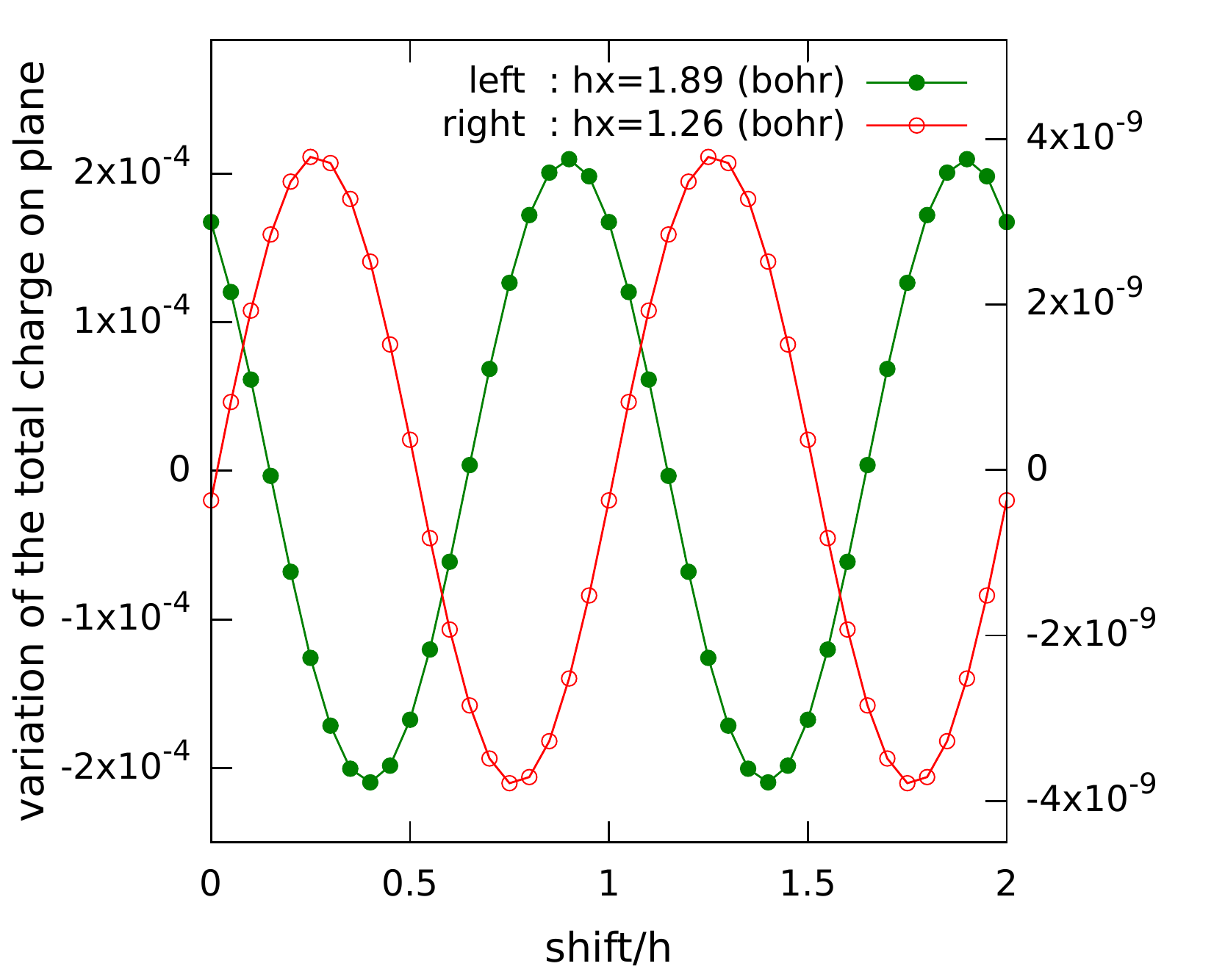}
\caption{(color online) Total induced charge on metallic plates versus shift of charged
particles in a periodic direction for two different values of grid spacing.
\label{fig:qerr}}
\end{figure}
Another interesting quantity in electrostatic problems with metallic boundaries
is the surface charge density induced on boundaries.
In order to investigate the accuracy of the method for
the surface charge density, we assess the total charge induced on the boundaries.
Due to translational invariance of the system in $x$ and $y$ directions,
the total charge induced on upper and lower planes must remain constant
under translation of particles along these directions.
In fact, the total charge oscillates due to the discretization of equations.
Fig.~\ref{fig:qerr} shows the oscillation amplitude of the total charge
induced on planes with respect to the translation in periodic directions for two values of
grid spacings in the periodic directions.
The deviation from constant value is very small.
Furthermore, the oscillation amplitude of the total charge decays rapidly
by more than five orders of magnitude as the grid spacing is reduced
by about $30$ percent.

\begin{figure}
\centering
\includegraphics[width=0.45\textwidth]{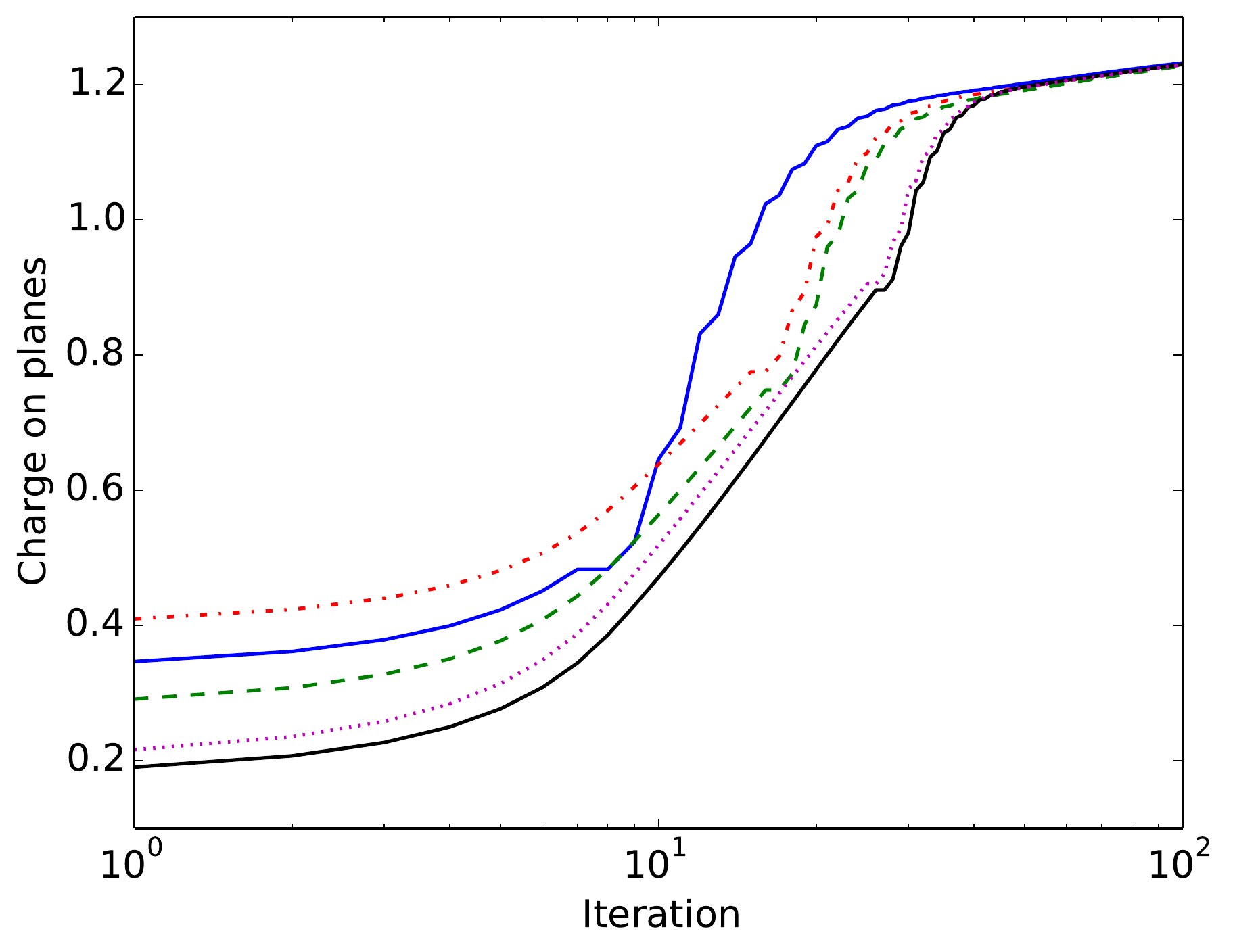}
\caption{(color online) Total charge induced on the metallic plates versus
iteration during minimization process.
Each curve represents different sodium chloride initial configuration in which
particles are slightly randomly displaced from the rock salt structure.
The final value of induced charge is independent of the trajectory since the
final structure of all different initial configurtions is the same.
\label{fig:q_iter}}
\end{figure}

We performed zero temperature atomistic simulation of sodium chloride
system containing $1000$ particles using structural relaxation
by means of energy minimization.
The short range interactions are obtained from the 
Born-Mayer-Huggins-Fumi-Tosi~\cite{Tosi1964} (BMHFT) rigid-ion potential,
with the parameters given in Ref.[\onlinecite{Ree1973}].
Fig.~\ref{fig:q_iter} illustrates the total charge induced on the upper plane
during minimization process of several structures with initial
configurations chosen to be randomly displaced particles from equilibrium.
As it is shown in Fig.~\ref{fig:q_iter}, the final configurations
have the same total charge induced on upper/lower planes
even though the trajectories are different and
induced charges are different during various trajectories.
Obviously, the total charge induced on the upper plane for the final configuration
in a minimization process could be different from others
if the final configuration would be different from those of other minimization processes.
This could happen if
the random displacement of the initial structures were too large.

\begin{figure}
\centering
\includegraphics[width=0.45\textwidth]{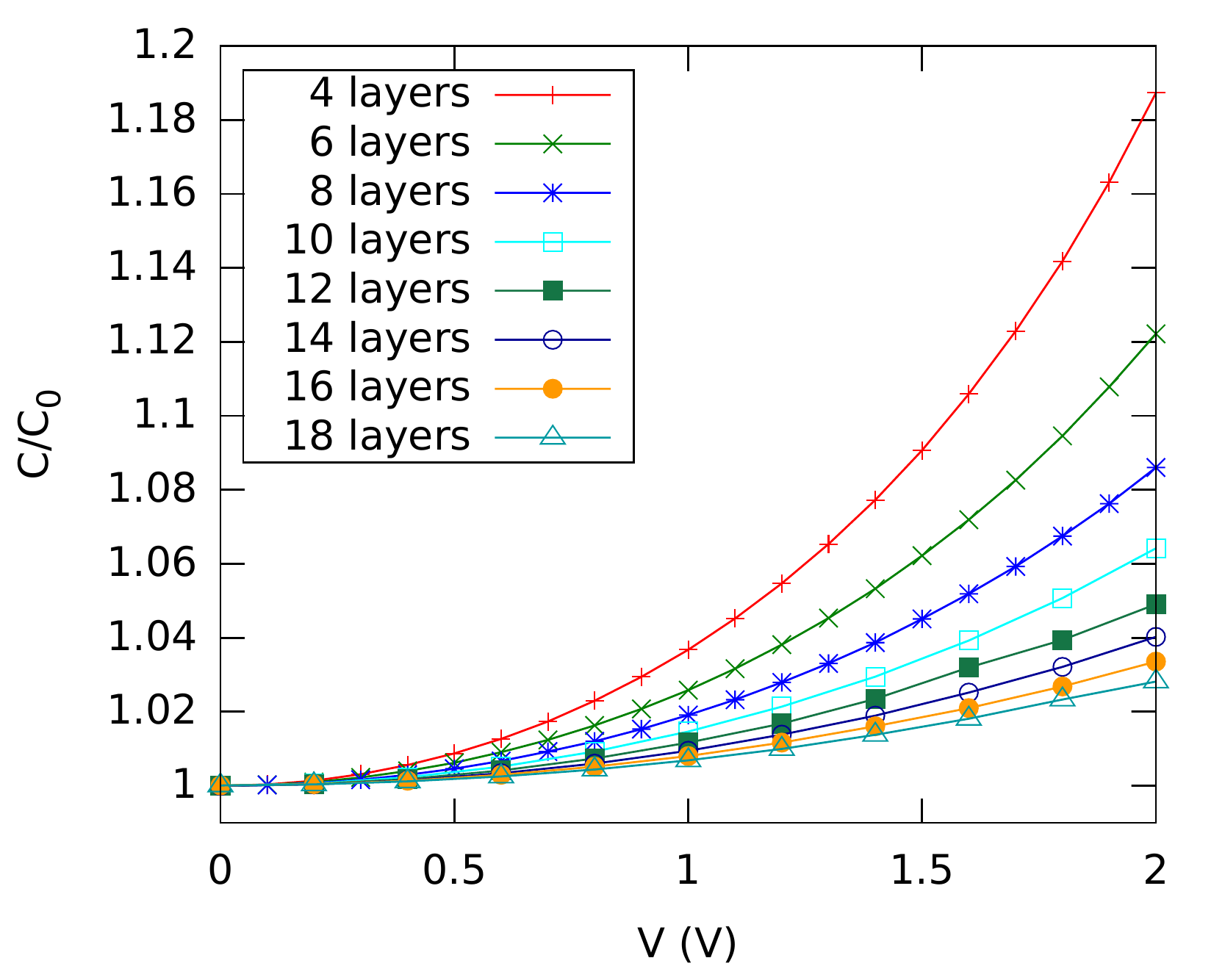}
\caption{(color online) Capacitance normalized to capacitance at zero bias voltage
versus potential difference of the two metallic parallel plates.
($ C_0 = \frac{dQ}{dV}|_{_{V=0}} $)
\label{fig:rel_c_v}}
\end{figure}

\begin{figure}
\centering
\includegraphics[width=0.45\textwidth]{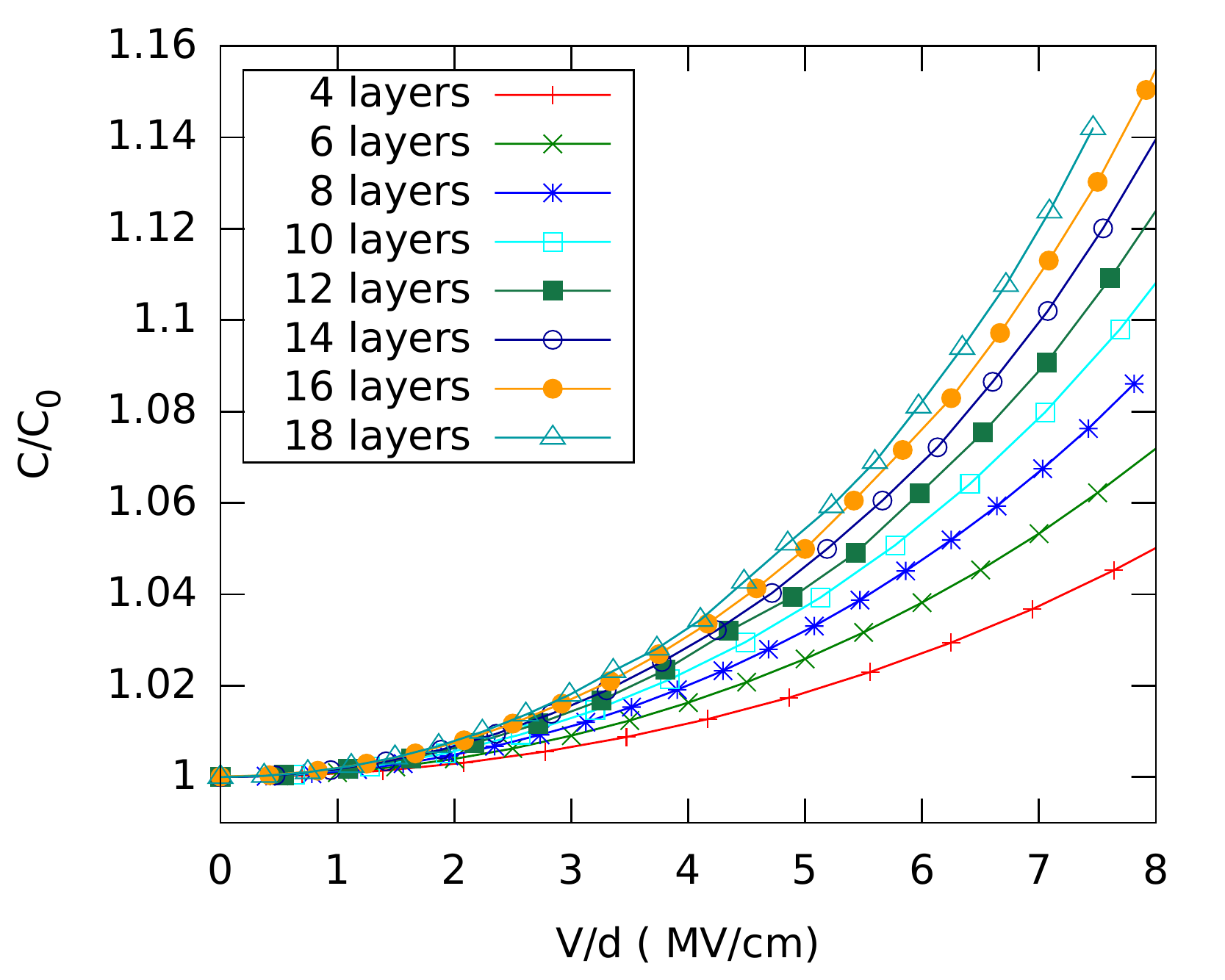}
\caption{(color online) Capacitance normalized to capacitance at zero bias voltage
versus average external electric field due to the two metallic parallel plates.
\label{fig:rel_c_e}}
\end{figure}

The response of insulators and semiconductors to external electric field
is of great interest both from experimental and theoretical aspects.
The typical approach to study external electric field is
to apply an external uniform electric field, however,
in experimental setup such electric field is caused by
voltage difference of two electrodes surrounding the material.
In the uniform electric field approach one cannot study
nonlinear response of the material to the external electric field.
This is of great significance for ultrathin films.
Furthermore, in practical situations ions do not feel a homogeneous
electric field and thus ionic displacements are indeed nonuniform.
Ions in the vicinity of the interface of the material and the electrodes
feel a more oscillatory electric field and studying this phenomena
is possible with the approach presented in this paper.
Here we study nonlinear dielectric behavior of sodium chloride
as a prototype of alkali halide.
The nonlinear behavior shows up in large electric fields.
Here we investigate this phenomena for ultrathin NaCl films
from $4$ to $18$ layers.
Due to the nonlinear behavior in the response of the material to
electrostatic field, the total charge induced on upper/lower plane is
not a linear function of the bias voltage, consequently,
the capacitance is not a constant and it should be calculated by
\begin{align*}
C(V)=\frac{dQ(V)}{dV}.
\end{align*}
Fig.~\ref{fig:rel_c_v} illustrates the ratio of capacitance to
the capacitance at zero bias limit in terms of bias voltage.
Fig.~\ref{fig:rel_c_v} shows that the nonlinear contribution to
capacitance increases strongly as a function of bias voltage and
it can be as much as $20\%$ of the linear contribution.
Based on the curves in Fig.~\ref{fig:rel_c_v},
one may assume that the thiner the film the larger is the
nonlinear contribution. 
This is in fact the opposite.
The nonlinear behavior is stronger for thicker films.
Fig.~\ref{fig:rel_c_e} shows the ratio of capacitance to
the capacitance at zero bias limit in terms of electric field.
The electric field is calculated by ratio of the potential difference
between the two electrodes to their distance.
From curves in Fig.~\ref{fig:rel_c_e}, it is obvious that the
nonlinear contribution to capacitance increases as the thickness
of the film grows for a given value of electric field.
However, the increase in the nonlinear contribution slows down
for thick films and it is expected to approach to zero
at the limit of very thick films.
Electric field in the calculations shown in Fig.~\ref{fig:rel_c_e},
are very large and cannot be applied to bulk NaCl since they are
larger than electrical breakdown of NaCl.
In general electrical breakdown of thin films is larger
than that of bulk and such large electric fields used
in these calculations may be plausible.

We have also calculated the dielectric constant of sodium chloride films
with respect to the thickness of the film.
Dielectric properties of sodium chloride films have been investigated
using ab initio techniques in Ref.~[\onlinecite{Ono2005}], where a uniform
external electric field is applied to sodium chloride films.
It would be very interesting if one could apply a combination of our method and
ab initio techniques such as density functional theory to investage dielectric properties of
ionic ultrathin films confined with metallic plates.
Table~\ref{tab:dielec_const} presents the list of dielectric constant for
NaCl ultrathin films with thickness of $10$ to $100$ layers.
The dielectric constants are obtained with relation $K=\frac{C_0}{C_{vac}}$
where $C_{vac}$ is the capacitance in the absence of the film.
$C_0$ is the capacitance at zero bias limit and for thick films it is virtually the same as
the value obtained by the ratio of the total charge induced on upper/lower plane
to the potential difference of the two electrodes.
A relaxation with very tight convergence criteria is required for thick films and
the system becomes ill conditioned with the growth of the number of particles.
Indeed, these two points prevent us to calculate the dielectric constant for very
thick films.
Table~\ref{tab:dielec_const} indicates that
the dielectric constant increases with the growth of the NaCl films,
however, the growth slows down for thick films.
It is expected that the diecltric constant approaches to a fixed value
which corresponds the dielctric constant of bulk sodium chloride.
In fact, the quaility of the interatomic potential is decisive for the
accuracy of the dielectric constant.
In Fumi-Tosi force field ions are treated with fixed charges and
therefore, the dielectric constant obtained in our approach accounts
only for the ionic degrees of freedom of the material and
does not include the effects due to the electronic polarization.
\begin{table}[t]
\caption{Dielectric constant ($K$) of various NaCl ultrathin films with
respect to the number of layers ($n$) in the film.
\label{tab:dielec_const} }
\begin{ruledtabular}
\begin{tabular}{l|ccccccccc}
$n$ &  10  & 20   & 30   & 40   & 50   & 60   & 80   & 100  \\ \hline
$K$ & 3.73 & 4.01 & 4.12 & 4.18 & 4.28 & 4.24 & 4.28 & 4.31 \\
\end{tabular}
\end{ruledtabular}
\end{table}

\section{Conclusion}
We presented a new method for electrostatic interaction of charged point particles
confined between two parallel metallic plates.
Due to the linear nature of governing electrostatic equations,
the electric potential is assumed to be a superposition of two parts;
one singular part in the absence of metallic plates that is due to the
existence of point charges in the simulation box, and the other is a smooth potential,
imposed by the metallic boundaries.
In this way, we were able to employ a previously developed method for the similar geometry
without the presence of the metallic plates.
The second part was solved in a similar approach but slightly different due to the resulting
homogeneous ordinary differential equation that could be solved analytically.
The method is very efficient and it has a quasilinear complexity of $\mathcal{O}(N\ln(N))$.
In order to investigate the accuracy and efficiency of the method, zero temperature atomistic 
simulations of sodium chloride system were performed.
The relative error of the total energy depends on the grid spacing, however,
it decreases very rapidly with reducing the mesh size and it can be
made sufficiently small, even zero limited by the machine precision.
Furthermore, by imposing the potential difference between metallic plates and allowing the system to relax,
nonlinear behavior of capacitance of sodium chloride ultrathin films was investigated.
The nonlinear behavior grows as the number of layers in the film is increased, however,
the growth slows down as the film thickness increases.
In addition, we calculated the dielectric constant of the sodium chloride ultrathin films
where it was shown that the dielectric constant increases with growth of the film.
The dielectric constant of sodium chloride films with $100$ layers is about $20\%$
lower than experimental dielectric constant of bulk sodium chloride.
In a forthcoming study we will employ our method and
investigate finite temperature effects on dielectric
properties of sodium chloride ultrathin films.

\clearpage
\providecommand{\noopsort}[1]{}\providecommand{\singleletter}[1]{#1}%

\end{document}